\def\itot{I_{\mbox{tot}}}
\def\ith{I_T}

\def\its{I_T^*}

\def\jc{j_c}
\def\jn{j_s}
\def\jtot{j_{\mbox{tot}}}

\def\et{E_T}

\def\ets{E_T^*}

\def\nc{n_c}
\def\nn{n_s}

\documentstyle[amssymb,aps,prl,epsf,multicol]{revtex}
\begin{document}
\tighten
\title{Temporally ordered collective creep and dynamic transition in the
charge-density-wave conductor NbSe$_3$
}

\author{S.~G.~Lemay and R.~E.~Thorne\\} 
\address{ Laboratory of Atomic and Solid State
Physics, Clark Hall, Cornell University, Ithaca,~NY~14853-2501\\ }
\author{Y.~Li and J.~D.~Brock} 
\address{ School of Applied \& Engineering Physics, Cornell University, 
Ithaca, NY 14853-2501 }
\date{\today} \maketitle

\begin{abstract} 

We have observed an unusual form of creep at low temperatures in the charge-density-wave (CDW) conductor NbSe$_3$. This creep develops when CDW motion becomes limited by thermally-activated phase advance past individual impurities, demonstrating the importance of local pinning and related short-length-scale dynamics. Unlike in vortex lattices, elastic collective dynamics on longer length scales results in temporally ordered motion and a finite threshold field. A first-order dynamic phase transition from creep to high-velocity sliding produces ``switching" in the velocity-field characteristic.  

\end{abstract}
\pacs{PACS numbers: 71.45.Lr, 72.15.Nj, 74.60.Ge}  

\vspace{-.6cm}

\begin{multicols}{2}
 \narrowtext    


Interaction between internal degrees of freedom and disorder
determines the dynamical properties of driven periodic media, a class
of systems that includes vortex lattices in type II superconductors 
\cite{fllreview}, Wigner crystals \cite{wigner}, magnetic bubble 
arrays \cite{bubbles}, and charge-density waves (CDWs), the
low-temperature phase of quasi-one-dimensional conductors
\cite{reviews}.  While the interplay between quenched
disorder and elastic deformations is relatively well
understood, a complete description of the role of thermal
disorder and plastic deformations, which may result in
disordered dynamical phases such as driven glass, smectic and liquid
states \cite{balents98}, has not yet been achieved.

CDWs have long been regarded as a prototypical system for the study of
many-degree-of-freedom dynamics,
both because of their relative theoretical simplicity and because
CDW materials like NbSe$_3$ exhibit collective 
phenomena with remarkable clarity.  A CDW consists of coupled modulations of
the electronic density $n=n_0+ n_1\cos[Q_c x+\phi(x)]$ and of the positions of the lattice ions
\cite{reviews}.  Applied electric fields $E$ greater than a threshold
field $\et$ cause the CDW to depin from impurities and slide relative
to the host lattice, resulting in a non-linear dc current density
$\jc$ proportional to the CDW's sliding velocity.  The impurities
cause the CDW to move nonuniformly in both space and time, and the elastic collective
dynamics leads to oscillations (``narrow-band noise") in $\jc(t)$.  The
frequency $\nu$ of these
oscillations is proportional to the dc component of $\jc$, and their
$Q=\nu/\Delta \nu$ can exceed 30,000 in high-quality NbSe$_3$
crystals.

\begin{figure}[hbt]
\epsfxsize=7.5 cm   
\centerline{\epsfbox{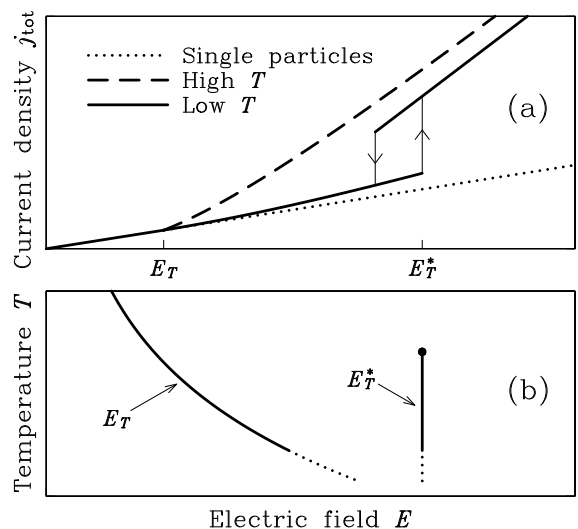}}
\vspace{.2cm}
\caption{
(a) Form of $\jtot(E)$ in the CDW conductor NbSe$_3$.  Dotted line: single particle current
density $j_s\propto E$.  Dashed line: total current density
$\jtot=j_s+\jc$ at high temperatures ($T>2T_P/3$).  Solid line: $\jtot$ at low $T$. The difference between
the solid or dashed lines and the dotted line gives the CDW current
density $\jc$.  (b) Temperature dependence of $E_T$ and $\ets$ in NbSe$_3$.
}
\label{fig:intro}
\end{figure}

Despite these simplifying features, most aspects of CDW transport at low temperatures remain poorly
understood. At temperatures $T>2T_P$/3, $\jc$ above $E_T$ is a smooth, asymptotically
linear function of $E$.  However, at low temperatures $\jc(E)$ changes drastically, as illustrated in
Fig.~\ref{fig:intro}.  CDW conduction still begins at $\et$ but $\jc$ is small and freezes out with decreasing temperature for fields less than a second characteristic field $\ets>\et$.  At $\ets$ $\jc$ increases by several orders of magnitude to a more nearly temperature-independent value, often by an abrupt, hysteretic "switch." Similar behavior
is observed in all widely studied CDW materials and is thus a
fundamental aspect of CDW dynamics
\cite{reviews,lowtrefs,itkis90zz93,hall,adelman93}.

\vspace{-.4cm}

We have characterized the sliding CDW's transport and
structural properties in the low-temperature regime of extremely high quality NbSe$_3$ crystals. For $\et<E<\ets$, we find that $\jc$ is activated in
temperature and increases exponentially with field.  Contrary to 
previous observations in driven periodic media, this
creep-like collective motion exhibits temporal order. Our results illuminate the relation between local and collective pinning and indicate that dynamics on lengths much shorter than the Fukuyama-Lee-Rice (FLR) length --- neglected in most theoretical treatments --- play a central role. They imply revised interpretations for ``switching" at $\ets$, the low-frequency dielectric response, low-field relaxation, and nearly every other aspect of the CDW response at low temperatures.


High purity ($r_R\approx 400$) whisker-like NbSe$_3$ single crystals with
typical cross-sectional dimensions of $\sim$3 by $\sim$0.8 $\mu$m were mounted
on arrays of 2 $\mu$m wide gold-topped chromium wires
\cite{adelman96}.  The total current density $j_{tot}$ is a sum of the
CDW and single-particle current densities, $\jc$ and $\jn$. $\jc$ is
orders of magnitude smaller than $\jn$ in the range $\et<E<\ets$
(except at relatively high temperatures and very close to $\ets$ \cite{adelman93}). $\jc(E)$ cannot be directly measured and its form in the low-velocity branch of NbSe$_3$ has not previously been determined.  As
shown in the inset to Figure 2, our high-quality crystals 
\cite{qualitynote} 
exhibit voltage oscillations with Q's as large as 130 in this regime.
Consequently,
we are able to determine $\jc$ by measuring the oscillation frequency
$\nu=(Q_c/2\pi e\nc)\jc$, where $-e$ is the electronic charge and
$\nc$ is the condensed carrier density.  
$\jc(E)$ was independently estimated by alternating the applied current's
direction and measuring resistance transients $R(t)$ associated with transients
in the
distribution of CDW strain $\epsilon(x)=(1/Q_c)(\partial\phi/\partial
x)$ between the current contacts \cite{adelman96,transnote}.



\begin{figure}[hbt]
\epsfxsize=7.5 cm   
\centerline{\epsfbox{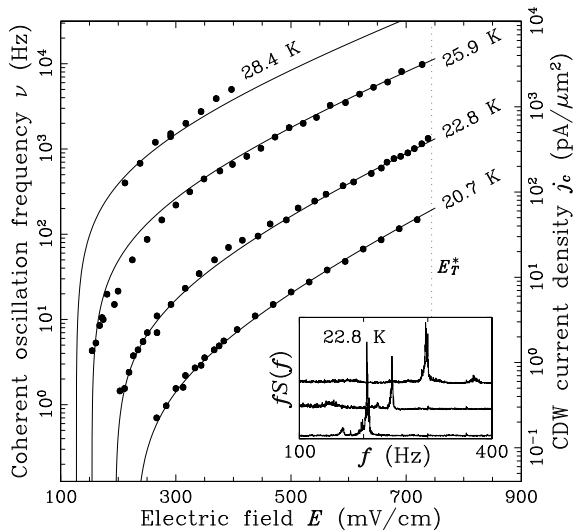}}
\vspace{.2cm}
\caption{
Coherent oscillation frequency $\nu$ and current density $\jc$ versus 
electric field $E$. The solid lines are a fit to Eq.~\ref{eq:fVfit}.
The intersection of the lines with the horizontal axis corresponds
roughly to the measured $\et$ at each temperature.  The dotted vertical line indicates $\ets$.  Inset: spectral
density $S(f)$ at 22.8~K for $E/\et$ = 2.63, 2.77 and 2.88; the curves
are offset vertically for clarity.  
}
\label{fig:fV}
\end{figure}

Figure~\ref{fig:fV} shows the CDW current density 
$\jc(E)=\nu\times0.32$~pA/$\mu$m$^2$Hz \cite{richard} calculated from the
measured oscillation frequency $\nu(E)$ for $\et < E
< \ets$ at four temperatures. The CDW moves extremely slowly
throughout this field and temperature range: the smallest measured
$\nu$ values at $T$=20.7~K correspond to CDW motion of roughly one wavelength
or 14 \AA\ per second and to $\jc\approx10^{-9} j_{tot}$. Between T$\approx$40
K and T$\approx$20 K, $\jc$ at fixed $E < \ets$ is temperature activated,
decreasing by roughly 7 orders of magnitude.   $\jc$
jumps abruptly at $\ets$, with
$\jc(E\!=\!1.1 \ets)/\jc(E\!=\!0.9 \ets)$ increasing from $\sim$ 10$^3$ to
$\sim$ 10$^6$ as $T$ decreases from 28 to 20~K. 


The current density $\jc \propto \nu$ can be fit by a modified form for thermal
creep \cite{anderson}
%
\begin{equation}
\jc(E,T)=\sigma_0 [E-\et] 
\,\exp \left[ -\frac{T_0}{T}\right] 
\,\exp \left[\alpha \frac{E}{T} 
\right].
\label{eq:fVfit}
\end{equation}
where the $[E-E_T]$ term describes the fact that the current drops to zero at a threshold $E_T$ that remains large even
at high temperatures. The solid lines in Fig.~\ref{fig:fV} indicate a fit with
$T_0=505$~K, $\alpha=136$ K\,V$^{-1}$\,cm, and $\sigma_0=350$ $\Omega^{-1}\mu$m$^{-1}$.
The value of $T_0$ is insensitive to the assumed field dependence and
corresponds to 0.6 times the single-particle gap $2\Delta$ \cite{tunnel},
consistent 
with measurements of delayed conduction \cite{levy92_93} and of
$\sigma_c$ near $\ets$ above 30~K \cite{adelman93}. Although creep is observed in other systems, the coherent oscillations imply that the creep in this case is highly unusual: it exhibits temporal order.

\begin{figure}[hbt]
\epsfxsize=7.5 cm   
\centerline{\epsfbox{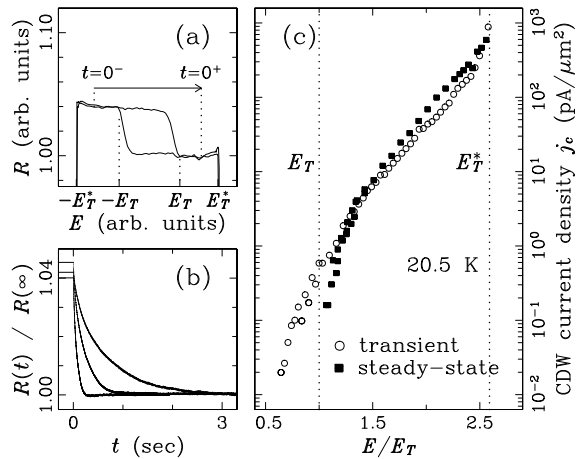}}
\vspace{.2cm}
\caption{
(a) Single-particle resistance $R$ of a 70 $\mu$m segment adjacent to a current contact versus electric field $E$. (b) $R(t)/R(\infty)$ for the same segment following a reversal of the polarity of $E$, as indicated by the arrow in (a), for $E/\et$=1.40, 1.69 and 1.93.  (c) Comparison of $\jc$ calculated from $R(t)$ \protect\cite{transnote} with $\jc$ obtained from measurements of the coherent oscillation frequency.  The current contacts were 630 $\mu$m apart, and $E_T$(20.5~K) = 49 mV/cm.
}
\label{fig:transient}
\end{figure}

Figure~\ref{fig:transient} shows $\jc(E)$ at $T$=20.5~K obtained
from transient measurements \cite{transnote}.  These data
agree closely with $\jc(E)$ deduced from $\nu(E)$ over
nearly three decades in $\jc$ \cite{comparenote}.  Combining the two
measurements yields $\jc/\nu=0.22$ pA/$\mu$m$^2$Hz, consistent with
the expected value of $\jc/\nu=0.32$ pA/$\mu$m$^2$Hz \cite{richard} within the
factor-of-two uncertainty in the value of $\jc$ determined from
transient measurements.  This rules out significant filamentary
conduction, observed in the low temperature regime of other CDW
materials,
and implies that the entire crystal cross
section or at least a significant fraction of it is involved in
coherent conduction.

\begin{figure}[hbt]
\epsfxsize=7.5 cm   
\centerline{\epsfbox{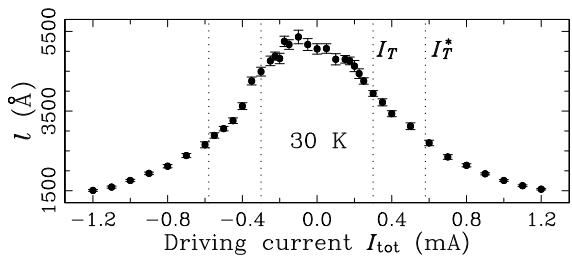}}
\vspace{.2cm}
\caption{
Inverse CDW peak half width (corrected for instrumental resolution) in
the [1~0~0] direction versus $\itot$.  $\its$ and an upper bound for $\ith$ 
were determined from measurements of $dV/d\itot$ and of the sharp increase in
1/f-like noise, respectively.
}
\label{fig:xray}
\end{figure}

Figure~\ref{fig:xray} shows the results of high-resolution x-ray diffraction measurements of the
CDW's transverse structure versus electric field. The CDW creates
superlattice peaks in
the diffraction pattern, and the half-width of each peak is inversely related
to the CDW phase-phase correlation length. For $E<\et$,
the resolution-corrected inverse half-width is $l \approx 4100$ \AA, comparable to the
crystal dimension in
this direction.  For $E>\et$, $l$ decreases monotonically with
increasing $E$, remaining greater than $2500$~\AA\ for $\et<E<\ets$. 
$l$ does not show any abrupt change at $\ets$ despite the several
orders-of-magnitude increase in $\jc$ there.  Similar results were obtained in other directions perpendicular to ${\bf Q_c}$ (e.g. [1~0~3] and [1~0~\={2}]) and at higher temperatures. 

Several different models have been proposed to account for the low-temperature
properties of CDW conductors. In K$_{0.3}$MoO$_3$ and TaS$_3$, whose Fermi
surfaces are completely gapped by CDW formation, the activation energies for the single-particle conductivity $\sigma_s$ and the
CDW conductivity $\sigma_c$ in the low-velocity branch are both comparable to the CDW gap so that $\sigma_c (T) \propto \sigma_s
(T)$ \cite{lowtrefs}.  Motivated by this observation, Littlewood
\cite{littlewood88} suggested that dissipation caused by single-particle screening of CDW deformations limits CDW motion in the low-velocity branch, and that an abrupt, hysteretic transition to the high-velocity branch occurs at a frequency $\nu$ comparable to the dielectric relaxation frequency $\nu_1\propto\sigma_s$ when this screening becomes ineffective. The predicted value of
$\nu$ at the discontinuity for K$_{0.3}$MoO$_3$ and TaS$_3$ is four orders of
magnitude too large \cite{adelman93}, and for NbSe$_3$ at $T$=20.7~K $\nu$ is 13
orders of magnitude too large.  Levy \textit{et al.} \cite{levy92_93} showed
that a related model exhibits a hysteretic transition from the pinned state to
a fast sliding state when $\sigma_s$ is small even if high-frequency screening
effects are neglected. Neither model can explain the low-temperature CDW
properties of partially-gapped NbSe$_3$, for which $\sigma_s$ remains metallic
and \textit{increases} with decreasing temperature below 50 K. 

Various forms of CDW plasticity including phase slip at isolated defects \cite{itkis90zz93,hall} and shear between two-dimensional CDW sheets
\cite{vinokur97} have been suggested to account for the properties of the
low-velocity branch and the transition at $\ets$. Our observation of highly
coherent oscillations in high-quality crystals and earlier results\cite{adelman93} rule out models based on slip at rare isolated defects and contacts, and our x-ray measurements
rule out the form of shear plasticity discussed in Ref.~\cite{vinokur97}. 

Brazovskii and Larkin\cite{brazovskii} have focused on the CDW's local 
interaction with defects. At low temperatures CDW phase advance past rare defects occurs
via thermally-activated  soliton generation, and motion becomes much more rapid at large fields when the effective barrier to soliton generation vanishes. This interpretation
has appealing features, but the suggested form for the $\jc(E)$ relation at low
temperatures does not reproduce the two branches separated by an abrupt
hysteretic transition or the field dependence in either branch observed
experimentally in NbSe$_3$.  Furthermore, the predicted $\ets$ is determined
by the soliton energy and should be independent of crystal size.
Experimentally, in NbSe$_3$ both $\et$ and $\ets$ vary as $1/t$ for crystal
thicknesses $t$ less than $\sim$ 20 $\mu$m \cite{adelman93}. The thickness
dependence of $\et$ results because transverse CDW correlations are limited by
$t$ so that collective pinning is two-dimensional \cite{mccarten}. 
Consequently, the thickness dependence of $\ets$ implies that it, too, is
determined by collective effects. 

CDW creep of a fundamentally different character is observed in thin NbSe$_3$ crystals at
high temperatures \cite{mccarten,creeprefs}.  Near T$_P$, $E_T$ is rounded, nonlinear
conduction can extend to near E=0, and highly coherent oscillations below the
nominal $E_T$ are not observed.  This incoherent creep occurs when $k_B T$
approaches the collective pinning energy ($\propto \Delta(T)^2 t$) of the phase-correlated FLR domains, which
in NbSe$_3$ have micrometer dimensions. The temporally-ordered creep observed in relatively thick crystals
at low temperatures above a sharp threshold $E_T$ must involve barriers that
are much smaller than those of collective pinning and that are not rare, and a length scale that is much smaller than
that of the collective dynamics responsible for the narrow-band noise.  

Motivated by earlier ideas \cite{brazovskii,abe,tutto}, we
suggest that the low-velocity branch develops when CDW motion becomes limited by 
thermally-activated phase advance by $\sim$ 2$\pi$ past individual
impurities. Although collective pinning is weak \cite{mccarten}, the phase of the $Q_c=2k_F$ oscillations is fixed at each impurity so that phase advance requires CDW amplitude collapse and a finite barrier $\sim \Delta$ \cite{tutto}. Collective dynamics within volumes containing enormous numbers of impurities (set by the FLR length) then generates the finite threshold $E_T$ and coherent oscillations, as in the high-temperature regime.  Unlike in vortex lattices, long-length-scale CDW dynamics is largely elastic and thus retains temporal order even though the short-length-scale dynamics is stochastic. 

The $E-E_T$ prefactor \cite{eqexpl} and the remaining terms in
Eq.~\ref{eq:fVfit} follow naturally from this combination of long and short length-scale processes. The measured barrier $T_0$ is consistent with the
expected pinning barrier per impurity of $\sim$ $\Delta$ \cite{tutto}. An
applied electric field should reduce this barrier by $\sim$ $e n_c V \lambda
E$, and using a condensate density $n_c = 2 \times 10^{21}$ cm$^{-3}$ \cite{reviews,mccarten}, a CDW
wavelength $\lambda=14 ~\AA$ and the measured $\alpha$ value yields a
volume $V$ involved in each thermally-activated event of $V \approx 4.2 \times
10^{-17}$ cm$^{3}$.  Using the scale factor expected for typical impurities
\cite{mccarten,impscale}, the bulk residual resistance ratio of $\sim$ 400 for our
crystals corresponds to a concentration $n_i \approx 2.5 \times 10^{16}$ cm$^{-3}$. 
The volume per impurity $1/n_i \approx 4 \times 10^{-17}$ cm$^{3}$ is thus
in excellent agreement with $V$ deduced from creep  measurements.

The present experiments together with those of Ref.~\cite{adelman93} rule out all previous explanations of the ``switching" between low and high velocity branches at $\ets$ in NbSe$_3$.    We suggest that switching occurs via a first-order dynamic phase transition\cite{balents98,koshelev94}.  The long-length-scale dynamics exhibits temporal order in both branches, but in the high-velocity branch dynamic fluctuations produced as the CDW moves past impurities may become more important than thermal fluctuations in overcoming impurity barriers \cite{koshelev94}. The transition's abruptness, hysteresis, and temperature dependence
shown in Fig.~\ref{fig:intro}, the CDW's tendency to fragment near the transition into distinct conducting regions \cite{hall}, and the field and temperature-dependent time delays required for the transition's completion \cite{levy92_93} are all consistent with this explanation. 

Finally, we note that local temporal order has very recently been observed in the creep regime of a vortex lattice \cite{troy}.


We thank C.L. Henley, M. B. Weissman, M. C. Marchetti, A. A. Middleton, and S. Brazovskii for fruitful discussions.  This work was
supported by NSF Grants DMR97-05433 and DMR-98-01792.  S.G.L. acknowledges
additional support from NSERC.  The x-ray data were collected using
beam line X20A at the National Synchrotron Light Source (NSLS).
Sample holders were prepared at the Cornell Nanofabrication
Facility.


\end{multicols}{}

\begin{references}
\bibitem{transnote}
As discussed in Ref.~\cite{adelman96}, $\jc$ at the midpoint between current
contacts just after the current is reversed is given by
%
$\jc(t\!=\!0^+) \, = \,(e\nc b/\eta)\,\partial/\partial t
[R(t)/\bar{R}]_{t=0^+}$,
where $b$ is the length of a short sample segment bounded at one
end by a current contact, $R(t)$ is its resistance, $\bar{R}$ is the
steady-state value of $R$, $\eta\approx\nc/\nn\approx200$, and $\nn$ is
the single-particle density.
\bibitem{comparenote}
The discrepancy at low $E$ is due to the elastic force associated with
the strain profile \cite{adelman96}.
\bibitem{impscale} This assumes $r_R n_i \approx 1 \times 10^{19}$ cm$^{-3}$,  measured for Ti impurities in Ref.~\cite{mccarten}, as an estimate for the nonisolectronic defects that determine $r_R$ in our high-purity crystals.
\end{references}
\end{document}